\documentstyle[12pt,epsf]{article}
 
 \newcommand {\be} {\begin{equation}}
\newcommand {\bea} {\begin{eqnarray} \nonumber }
\newcommand {\ee} {\end{equation}}
\newcommand {\eea} {\end{eqnarray}}
 \newcommand {\eps} {\epsilon}

\newcommand {\ba} {\overline}

\def \form#1 {eq. (\ref{#1}) }
\def \parziale#1#2  {{\partial {#1} \over \partial {#2}}}

\begin{document}

\title{On the approach to equilibrium  of an Hamiltonian chain of anharmonic oscillators}
\author{
Giorgio
Parisi\\
Dipartimento di Fisica, Universit\`a {\em La  Sapienza},\\ 
INFN Sezione di Roma I \\
Piazzale Aldo Moro, Roma 00185, Italia}
\maketitle

\begin{abstract}
In this note we study the approach to equilibrium of a chain of anharmonic oscillators.  We
find indications that a sufficiently large system always relaxes to the usual equilibrium 
distribution.  There is no sign of an ergodicity threshold.  The time however to arrive to 
equilibrium diverges when $g \to 0$, $g$ being the anharmonicity.

\end{abstract}
A  debated issue is the approach to equilibrium of an Hamiltonian system.  A well studied 
problem is a chain of anharmonic oscillators.  The first numerical simulations have been done more 
than 40 years ago \cite{PASTA}: the authors found that for sufficiently small anharmonicity the 
system does not goes to the usual Boltzmann Gibbs equilibrium and there is strong memory of the 
initial conditions especially if the systems starts from a relatively smooth configurations, i.e.  
only long wave length modes are populated.

In the case of a finite system the KAM theorem \cite{KAM} states that for small anharmonicity $g$ 
the behaviour of the system is not ergodic and there is threshold value of the anharmonicity at 
which the system becomes ergodic.  Unfortunately the KAM theorem cannot be applied in the limit of 
infinite volume systems and no conclusions can be obtained from the KAM theorem in the case of a
very long chain.

  This is not a {\em technical} problem: when the volume becomes infinite  the
spectrum of the frequencies of the harmonic Hamiltonian becomes continuous. This phenomenon destroys 
the no-resonance condition which is at the heart of the KAM theorem.  Indeed naive arguments 
 would suggest that for small anharmonicity $g$ the time $\tau$ needed to reach equilibrium
diverges  as $g^{-2}$.  It was also argued 
\cite{TANTI,parisibook}that the the system is always reaching equilibrium and that a simple
physical argument implies that the time $\tau$ cannot diverge at small $g$ faster than $\exp(A/g)$.

A careful study of the original Fermi-Pasta-Ulam system can be found in \cite{KLR,DLR,Poggi,FI}.  
In this case the degrees of freedom of our system are $N$ variables $q_{i}$ and $p_{i}$.  The 
Hamiltonian is
\be
H=\sum_{i=1,N}\left({p_{i}^{2}\over 2}+  +{(q_{i}-q_{i+1})^{2}\over 2 }
+g {(q_{i}-q_{i+1})^{4}\over 4}\right),
\ee
where we impose periodic boundary conditions (i.e.  $N+1\equiv 1$).  Indications for the absence of 
an ergodicity threshold and for a divergent equilibration time where found in \cite{FI}.

In this note we have studied a different Hamiltonian, i.e. 
\be
H=\sum_{i=1,N}\left({p_{i}^{2}\over 2}+{q_{i}^{2}\over 2} +{(q_{i}-q_{i+1})^{2}\over 2 }
+g {q_{i}^{4}\over 4}\right).
\ee

In the first case the harmonic Hamiltonian (i.e.  in the case where $g=0$) contains all the 
frequencies squared which go from 0 to 2, while in the second case the frequencies squared of the 
harmonic Hamiltonian go from 1 to 3.  Moreover if we populate only the modes with Fourier transform 
concentrated at small momenta, in the first case the range is the interval $[0,\eps]$ in the second 
$[1,1+\eps]$, where $\eps$ is of the order of the square of the maximum momentum.  In other words in 
the first case the spectrum of excitation is always wide, while in the second case is quite narrow.  
This difference in the form of the energy spectrum may have drastic implications on the speed of the 
approach to equilibrium.

In this not wee have done a careful analysis of the numerical simulations, paying a particular 
attention to the choice of the initial condition.

The initial conditions we use are
\bea
p_{i}=\sum_{k=1,N/16}\left( a_{k} \cos ({2 \pi k i \over N}) + b_{k} \sin ({2 \pi k i \over 
N})\right)\\
x_{i}=\sum_{k=1,N/16}\left( c_{k} \cos ({2 \pi k i \over N}) + d_{k} \sin ({2 \pi k i \over 
N})\right)
\eea
In this way only Fourier modes for $k \le N/16$ are different from zero at time $t=0$.  For $g=0$ 
this property will remain valid at all time.

The solution of the Hamiltonian
equations (for $g=0$) is
\bea
p(t)_{i}=\sum_{k=1,N/16}\left( a(t)_{k} \cos ({2 \pi k i \over N}) + b(t)_{k} \sin ({2 \pi k i \over 
N})\right)\\
x(t)_{i}=\sum_{k=1,N/16}\left( c(t)_{k} \cos ({2 \pi k i \over N}) + d(t)_{k} \sin ({2 \pi k i \over 
N})\right),
\eea
where
\bea
a(t)_k=a_k \cos(\omega(k)t)+c_k \omega(k)^{-1} \sin(\omega(k)t), \ \ \ 
c(t)_k=c_k \cos(\omega(k)t)+a_k \omega(k)\sin(\omega(k)t), \\
b(t)_k=b_k \cos(\omega(k)t)+d_k \omega(k)^{-1} \sin(\omega(k)t), \ \ \ 
c(t)_k=d_k \cos(\omega(k)t)+b_k \omega(k)\sin (\omega(k)t),
\eea
where
\be
\omega(k)=(1+(2 \pi k / N)^{2})^{-1/2}\label{DISP}
\ee

The variables $a, b, c$ and $d$ can be chosen randomly at the initial time.  The advantage of a 
random choice it to allow an analytic computation.  Moreover in the case of a random choice we can 
perform an ensemble average which dumps the oscillations which are present for any particular choice 
of the initial condition.  In this way, after the average, we obtain a smoother dependance on the 
time.  

In this note we consider the ensemble
\bea
a_{k}=\frac{1}{2} \cos(\phi_{k})\ \ \ b_{k}=\frac {1}{2} \sin(\phi_{k}),\\
c_{k}=\frac{\omega(k)}{2} \cos(\psi_{k})\ \ \ c_{k}=\frac{\omega(k)}{2} \sin(\psi_{k}),
\label{INIT}
\eea
where $\phi_{k}$ and $\psi_{k}$ are random variables uniformly distributed in the interval $0-2 \pi$.
\begin{figure}[htbp]
\epsfxsize=400pt\epsffile[22 206 549 549]{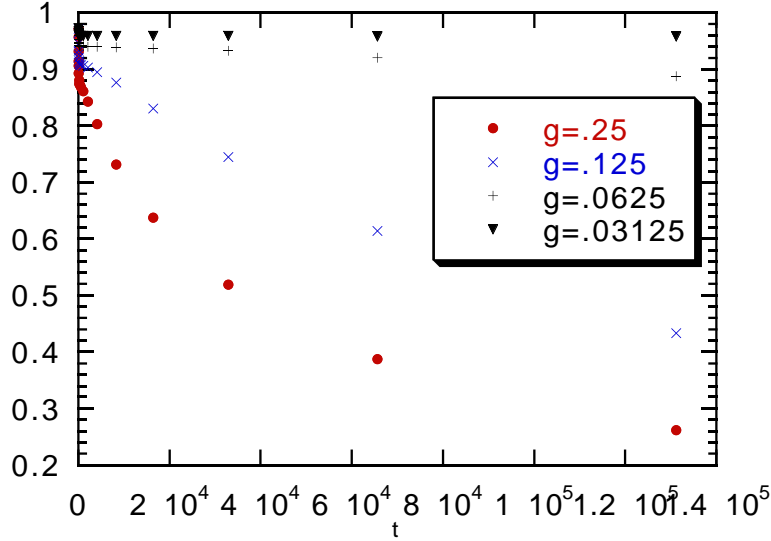}
\caption{ The function $\Delta(t)$ as function of the time $t$  for various values
of
$g$
(i.e.
.25,.125
 .0625, 03125).}
\end{figure}

In the infinite volume limit this ensemble (as far local observables are concerned) is equivalent to
the one in which variables $a, b, c$ and  $d$ are uncorrelated random variables with variance
\be
<a_{k}^{2}>=<b_{k}^{2}>=\frac12 \ \ <c_{k}^{2}>=<d_{k}^{2}>=\frac{A_{k}^{2}}2
\ee

The Gaussian ensemble has the advantage to be invariant under the time evolution when $g=0$. We
have used the ensemble define in eq. (\ref{INIT}) in order to start from a system with fixed value of 
the total energy in the limit $g=0$.  In this way we  suppress fluctuations of the total energy 
present in the Gaussian ensemble which are potentially annoying especially for small system.

The aim of this note is to study the time dependence of the ensemble average of various local 
observables in the case of a large system.  Here we report results for $N=8192$, simulations at 
smaller value of $N$ (i.e.  $N=2048$ and $N=16384$ do not differ significantly).  The time evolution 
was done by integrating (in double precision) the Hamiltonian equations of motion with a small time 
step  with the leap frog method (we have done simulations with  $\delta t= 0.05$ and $\delta t=.025$
and we have found no significant differences).  All the averages are done on an ensemble  of ten
different initial conditions.

\begin{figure}[htbp]
\epsfxsize=400pt\epsffile[22 206 549 549]{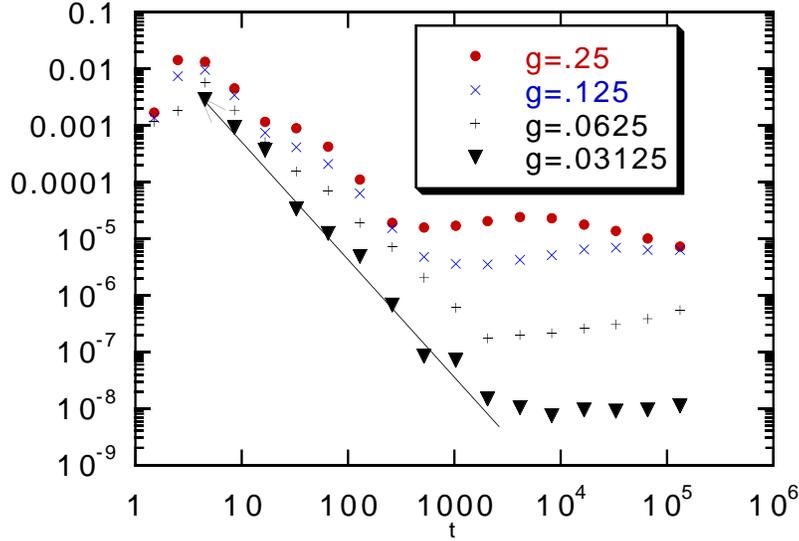}
\caption{The function $D(t)$ as function of the time $t$  for various values
of $g$ (i.e. .25,.125 .0625, 03125) on a double logarithmic scale.  The line is proportional to 
$1/t^{2}$.}
\end{figure}

The main quantity on which we concentrated our attention is $\Delta$, defined as
\be
\Delta(t)= {\ba{\sum_{i}q_{i}q_{i+1}} \over\ba{\sum_{i}q_{i}^2} }
\ee

We have chosen  the quantity $\Delta(t)$ for the following reasons:
\begin{itemize}
\item It is an intensive quantity which can be measured with high precision.
\item At $g=0$ it does not depend on $t$.
\item It starts from an high value ($\approx.97$) at $t=0$ and it must be zero in the Boltzmann 
Gibbs ensemble.  It value is therefore a very good indicator of equilibration.
\end{itemize}

The dependence of $\Delta(t)$ on $t$ is shown if fig.(1) for some values of $g$. 

We notice that the data at $g=.25$ are very well fitted at times larger then $10^{3}$ by a stretched 
exponential, i.e.  $\Delta(t) \propto \exp (- (t/\tau_{exp})^{1/2)}$, with $\tau_{exp}\approx .7\ 
10^{5}$.  Decreasing the value of $g$ the stretched exponential regime sets in at larger and larger 
valuer of time.
 
 The first 
impression would be that there is an ergodicity threshold.  i.e for $g>g_{T}$ $\Delta(t)$ goes to 
zero at $t$ infinity ,
while for $g<g_{t}$ $\Delta(t)$ goes to 
a non zero value at $t$ infinity.   The value of $g_{T}$ could be naively estimated 
around $.05$.

\begin{figure}[htbp]
\epsfxsize=400pt\epsffile[22 206 549 549]{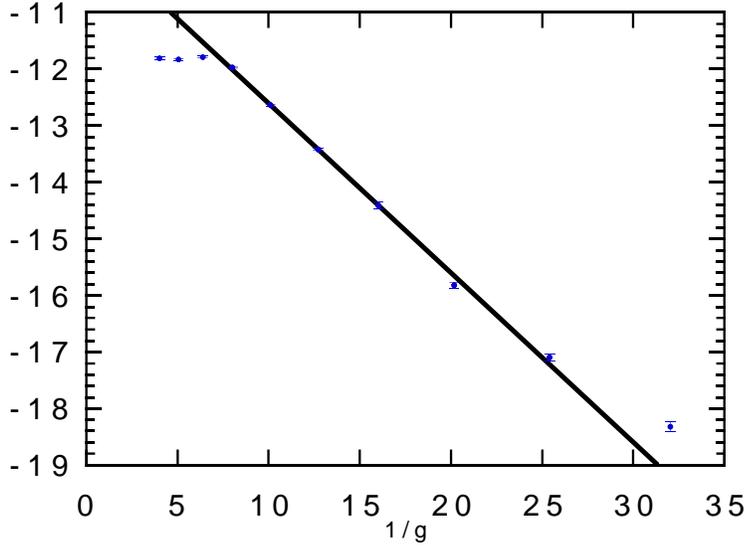}
\caption{ We plot of $\ln(\tilde\tau_{M}(g)))$ versus $g^{-1}$.  The line is a linear plot with slope
$.3$.}
\end{figure}

The presentation of the data is misleading and this impression is wrong.  We show in fig.(2) the 
data for
\be
D(t)\equiv {\partial \log(\Delta(t)) \over \partial t}.
\ee

We see that there are two regimes, at short times (up to a few hundreds) $D(t)$ decay as a power of 
the time (i.e.  as $t^{-\alpha}$ with an exponent $\alpha$ near 1.5.  At larger values of time 
the dependance of $D(t)$ flattens, indicating roughly an exponential decay.  The data at very large 
times for $g=.25$ decrease again, by this happens in the region where $\Delta(t)$ is much smaller 
that 1, indicating a stretched exponential behaviour in the tail at very large times.  In any case 
it is quite clear that there is no threshold and that the behaviour for different values of $g$ is 
quite similar, the only difference is that the time scales are much larger when $g$ become smaller.

The time needed to equilibrate (i.e. $\tau(g)$) is essentially the inverse of the value of $D(t)$ in 
the region where it is roughly constant.  Given this rather complex behaviour of $D(t)$ this 
definition is slightly ambiguous.  We have firstly studied the $g$ dependence of the quantity
$\tau_{M}(g)$ defined as
\be
\tau_{M}(g)\equiv D(t)^{-1}|_{t=t_{M}} \label{DEF1}
\ee
where $t_{M}$ is the largest time in our simulations, i.e.  $t_{M}=2.5\ 10^{8}$.  Obviously the 
choice of $t_{M}$ will influence the dependence of $\tau_{M}(g)$ on $g$.

The $g$ dependence of the $\tau_{M}(g)$ give us a qualitative information on the $g$ dependance of 
the relaxation time.  In order to do something better we should able to compare the values of 
$D(t,g)$ at homologous values of the time.  This will be done later.  For the moment let us stick to 
the definition eq.  \ref{DEF1}.  The results for this definition of $\tau_{M}(g)$ are shown in fig.3 
on a logarithmic scale.

There is a region were the data are compatible with exponential behaviour i.e.  $\tau_{M} \propto 
\exp (-A/g)$ with $A$=.3.  The fit is good in an intermediate region, it does not work at large 
values of $g$ (as expected), however some deviations are observed at small values.  We have tried a 
power fits (i.e.  $\tau_{M} \propto g^{-\lambda}$; the value of the exponent is around 5, but the 
fits are not good.
\begin{figure}[htbp]
\epsfxsize=400pt\epsffile[22 206 549 549]{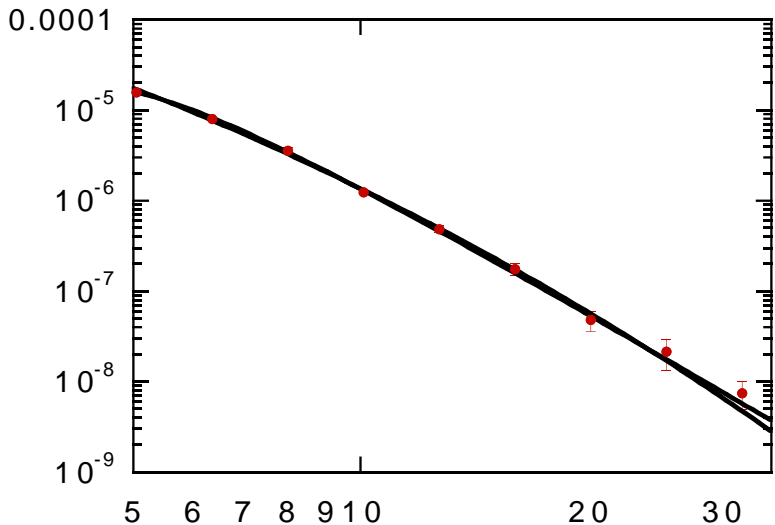}
\caption{ We plot  $\tau^{-1}$ versus $g^{-1}$, in a logarithmic scale where $\tau$ is  the
value of $D(t)$ at its first minimum.  The two curves are proportional to $\exp(-Ag^{\gamma})$, with 
$\gamma=.3$ and to $ag^{5}+b g^{6}$ respectively.}
\end{figure}

This analysis shows that this naive estimate of the correlation time does not show any sign of a 
threshold behaviour.  In order to estimate better the dependence of the equilibration time on $g$ we 
have to do something more systematic.  An obvious solution would to compute the time needed to reach 
a fixed value of $\Delta$ (e.g.  .5) or better to measure the time decay constant in the stretched 
exponential regime.  In order to implement such an approach one has to follow the system for a very 
large time.  Here we have followed a different strategy which can be implemented doing much shorter 
simulations.

A possible alternative definition of the correlation time could be done if we compare the curves for 
$D$ at homologous times.  We note that all the curves for $D$ have a minimum at a time which 
increases by decreasing $g$.  A different estimate of the equilibration time ($\tau(g)$) would be 
the function $D(t)$ evaluated at the minimum. 

Using this definition of $\tau(g)$ we plot the 
function $\tau(g)^{-1}$ versus $g$ in fig.  (4).  We can now fit the data in fig.  (4) both as 
$B\exp(-Ag^{\gamma})$ (with $B=1.6$,$A=7.1$ and $\gamma=.29$) or as $ag^{5}+b g^6$ (with $a=.22$ and 
$b=-.85$).

The two fits are equally good and we cannot distinguish among them.  They have rather different 
theoretical implications:
\begin{itemize}
\item
If the equilibration time diverges exponentially when $g \to 0$, this effect should be invisible in 
perturbation theory and it would be a non perturbative effects.  In this case we could define in the 
framework of perturbation theory a non trivial equilibrium state, which would become unstable as a 
consequence of non perturbative effects.
\item
A power like divergence of the correlation time (when $g \to 0$) is a perturbative effect which 
should be computable in the framework of perturbative expansion.
\end{itemize}

\begin{figure}[htbp]
\epsfxsize=400pt\epsffile[22 206 549 549]{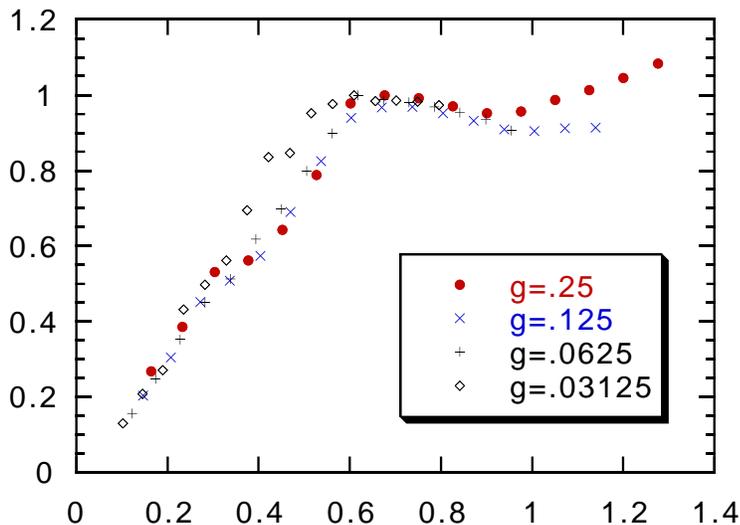}
\caption{ We plot of $\tilde\tau(g) \tilde D(t,g)$ versus $\ln(t)/\ln(\tilde \tau(g)$, in a 
logarithmic scale for for various values
of $g$ (i.e. .25,.125 .0625, 03125).}
\end{figure}

We have also explored if we can find some rescaling of the data for $D(t,g)$ from one value of $g$ 
to an other value of $g$.  In the short time region we expect that $D(t,g)=g f(t)$.  In order to 
take care of this dependence on $g$ we have defined
\be
\tilde D(t,g) =g^{-1}D(t,g).
\ee
In a similar way $\tilde \tau(g)$ is the value of $\tilde D(t,g)$ at the first minimum.  It would be 
interesting to see if some simple scaling law is satisfied.  We have tried with the scaling $\tilde 
D(t,g)=\tilde\tau(g)^{-1} f(t/\tilde\tau(g))$, but it does not work nicely.  We have tried a 
different scaling, i.e.  $\tilde D(t,g)=\tilde\tau(g)^{-1} f(\ln(t)/\ln(\tilde \tau(g)))$, and the 
results are much better, although the scaling is not perfect (see fig.  5).

We have also done numerical simulations in which we have used a different initial condition.  
Instead of eq.  \ref{DISP} we have used the modified form
\be
\omega(k)=(1+3g <x^{2}>+(2 \pi k / N)^{2})^{-1/2},
\ee
where $<x^{2}$ is determined in a self consistent way.  This initial condition takes care of the 
first order perturbative corrections to the probability distribution and it is characterized by a 
smaller value of $\Delta$ at short times.  The simulations have a quite similar behaviour.  One 
finds a very good power fit of the data for $\tau(g)$ of the form $A g^{-4.2}$.  The exponent for 
the time seems slightly different and this may be an effect of the way in which it has been defined.  
A more careful study is needed to obtain the precise $g$-dependence of the equilibration time.
It is interesting to recall that in the study of the original Fermi-Pasta-Ulam model a power law 
divergence of the equilibration time was found in \cite{FI}, although exponent is smaller (i.e.  
$\tau\propto g^{-3}$).

We have presented further evidence that a long chain of anharmonic oscillators always locally 
equilibrates in the infinite volume limit and that the equilibration time diverges in the limit of 
zero anharmonicity.  The precise form of this divergence is not fully determined.  More careful 
numerical experiments (also on other non linear equations) should be able to settle the question.

\section* {Acknowledgments}It is a pleasure for me to thank  S.  Franz and R.  Livi for useful
discussions.

\end{document}